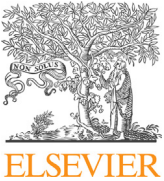
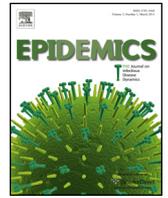

# Estimate of the reproduction number of the 2015 Zika virus outbreak in Barranquilla, Colombia, and estimation of the relative role of sexual transmission

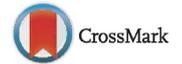

Sherry Towers [a,*], Fred Brauer [b], Carlos Castillo-Chavez [a], Andrew K.I. Falconar [c], Anuj Mubayi [a], Claudia M.E. Romero-Vivas [c]

[a] Arizona State University, Tempe, AZ, USA
[b] University of British Columbia, Vancouver, BC, Canada
[c] Universidad del Norte, Barranquilla, Atlantico, Colombia

## ARTICLE INFO

## ABSTRACT

*Background:* In 2015, the Zika arbovirus (ZIKV) began circulating in the Americas, rapidly expanding its global geographic range in explosive outbreaks. Unusual among mosquito-borne diseases, ZIKV has been shown to also be sexually transmitted, although sustained autochthonous transmission due to sexual transmission alone has not been observed, indicating the reproduction number (R0) for sexual transmission alone is less than 1. Critical to the assessment of outbreak risk, estimation of the potential attack rates, and assessment of control measures, are estimates of the basic reproduction number, R0.
*Methods:* We estimated the R0 of the 2015 ZIKV outbreak in Barranquilla, Colombia, through an analysis of the exponential rise in clinically identified ZIKV cases (n = 359 to the end of November, 2015).
*Findings:* The rate of exponential rise in cases was $\rho = 0.076 \, \text{days}^{-1}$, with 95% CI [0.066,0.087] $\text{days}^{-1}$. We used a vector-borne disease model with additional direct transmission to estimate the R0; assuming the R0 of sexual transmission alone is less than 1, we estimated the total R0 = 3.8 [2.4,5.6], and that the fraction of cases due to sexual transmission was 0.23 [0.01,0.47] with 95% confidence.
*Interpretation:* This is among the first estimates of R0 for a ZIKV outbreak in the Americas, and also among the first quantifications of the relative impact of sexual transmission.

## 1. Introduction

Zika virus (ZIKV), a mosquito-borne arbovirus, was first identified in Uganda in 1947. Similar to the dengue and chikungunya viruses, ZIKV is primarily spread by the tropical and sub-tropical domestic mosquito, *Aedes aegypti*. Recent outbreaks of ZIKV disease have occurred in French Polynesia in 2013-14 (Kucharski, 2016), and Yap Island in the Pacific in 2007 (Duffy et al., 2009), and since 2015, ZIKV has rapidly spread through many countries in South America, Central America, and the Caribbean where the *Aedes aegypti* vector species is endemic (Fauci and Morens, 2016).

ZIKV disease is usually asymptomatic, and is typically mild even upon clinical presentation, with symptoms similar to that of dengue virus (DENV) and chikungunya virus (CHIKV) infection (World Health, 2016; Fauci and Morens, 2016; Hayes, 2009). However, particular sets of clinical diagnostic criteria have been successfully employed, for example in Brazil (Brasil, 2016) and Colombia (Tolosa-Perez, 2016), to successfully differentiate patients with ZIKV infections from DENV and CHIKV infections (Brasil, 2016). ZIKV disease has been linked to an apparent increased risk of the neurological disorder Guillain-Barré syndrome, and also to neonate microcephaly (Fauci and Morens, 2016). The latter is of particular concern, because pregnant women may not know they have been infected, and the damage to their unborn infants may result in subsequent lifelong disabilities. In addition, there is evidence that there is also a direct human sexual transmission component to the disease, although it is unclear how significant a factor this aspect is in overall transmission in *Aedes aegypti* infested areas (World Health, 2015). Quantification of the role of ZIKV sexual transmission is cru-cial to the assessment of the relative efficacy of avoidance of sexual contact in case reduction.

* Corresponding author at: Simon A. Levin Mathematical, Computational and Modeling Sciences Center, Arizona State University, PO Box 873901, Tempe, AZ, 85287-3901, USA.
E-mail addresses: smtowers@asu.edu, sherrytowers@hotmail.com (S. Towers).

There is currently no vaccine or specific treatment for ZIKV infection (World Health, 2015), leaving control of the vector populations, use of mosquito repellents, and avoidance of sexual contact as the only means to control the spread of the disease.

Critical to the assessment of outbreak risk, and to the design, development, and evaluation of control strategies, are mathematical models that simulate the underlying dynamics of the transmission of a disease within a population (Hethcote, 2000). A particularly important quantity in infectious disease epidemiology that can be estimated with such models is the basic reproduction number, R0, which is the average number of secondary cases produced in a completely naïve population by the introduction of a single infectious individual (Hethcote, 2000; Heffernan et al., 2005). Very few estimates of the R0 of ZIKV disease have been published in the literature at the time of this writing (September, 2016), likely largely due to the surveillance challenges posed by this oftenasymptomatic disease in countries with limited human and capital resources for surveillance.

Here we examine the outbreak of ZIKV that began in 2015 in Barranquilla, Colombia. Barranquilla is the major Colombian port city on the Caribbean coast, with a population of 1.4 million (Romero-Vivas et al., 2013). It is a center of development in Colombia, and in recent years has experienced a rapid expansion of urbanization (Romero-Vivas et al., 2013). It has a tropical savannah climate, with an average daytime temperature of 32 °C year round, with day/night average of 28.4 °C. Wet seasons occur from April to June and August to November, with a yearly average precipitation of just over 800 mm, providing an ideal habitat for the *Aedes aegypti* mosquito (Romero-Vivas et al., 2006).

The public health department of the city has had a long-standing program of surveillance of arboviral diseases in both the human and vector populations; the dengue viruses have been endemic in the city for many years, with three serotypes co-circulating prior to 2007 (DENV-1, -2 and -4), and all four DENV serotypes circulating since then (Falconar et al., 2006; Falconar and Romero-Vivas, 2012). Since 2014, chikungunya virus also emerged in a major outbreak in Barranquilla, and affected a large fraction of the population (Cardona-Ospina et al., 2015).

In this work we used syndromic surveillance data, and examined the clinically identified cases of ZIKV in Barranquilla from the beginning of October to the end of December, 2015. We employed a mathematical model for vector-borne disease transmission with additional direct sexual transmission to assess the reproduction number, R0, for ZIKV disease spread, by fitting the parameters of the model to the initial exponential rise in cases. Given that sexual transmission alone has not resulted in sustained autochthonous transmission in areas free of the *Aedes* mosquito vector, we assume that the R0 corresponding to sexual transmission alone is less than 1. Based on this information, we used our model to assess the fraction of cases in Barranquilla generated by sexual transmission.

## 2. Methods and materials

### 2.1. Data

The data used consisted of the daily incidence of ZIKV disease cases identified during 2015 in Barranquilla, Colombia, by the syndromic surveillance system of the Colombian National System for the Public Health (Sistema de Vigilancia en Salud Publica: SIVIGILA) (Ministerio de la, 2016). Human ZIKV infections were clinically identified upon presentation at the multiple primary public health clinics within the city as distinct from those caused by DENV or CHIKV according to the National Health Institute national syn-dromic surveillance guidelines (Tolosa-Perez, 2016), as were also successfully employed in Brazil (Brasil, 2016), and other countries (see, for instance, (Tognarelli et al., 2015)).

**Table 1**
Number of cases of Zika virus disease in Barranquilla, Colombia, with date of identification between October 1, 2015 and December 31, 2015.

| Total | n = 1470 |
| --- | --- |
| Male | 610 (41%) |
| Non-pregnant female | 818 (56%) |
| Pregnant female | 42 (3%) |
| Age 0–17 | 232 (16%) |
| Age 18–44 | 901 (61%) |
| Age 45–64 | 289 (20%) |
| Age 65 and over | 48 (3%) |

A total of n = 1470 cases were identified between the beginning of October and the end of December 2015. The gender and age demographics of those cases are shown in Table 1. The dates of initial symptoms were available for n = 1141 (78%) of these cases. The incidence data, aggregated by day of initial symptoms, are shown in Fig. 1.

Of the n = 1470 cases identified in Barranquilla up to the end of December 2015, 3% occurred in pregnant women. Because concern over fetal birth defects may have prompted women to seek testing for ZIKV when they otherwise would not, there was the potential for surveillance bias. In the following, we thus determine the exponential rise in cases including pregnant women, and crosscheck the analysis by excluding pregnant women.

### 2.2. Statistical methods for fitting the exponential rise in incidence

In the early stages of an infectious disease outbreak, the number of incident cases grows exponentially in time as the effect of the increasing incidence on the depletion of the susceptible population remains small (Wallinga and Lipsitch, 2007).

We employed maximum likelihood methods to fit an exponential curve to the initial rise in daily ZIKV incidence data, by date of initial symptoms, to determine the initial exponential growth rate, $\rho$, using a Negative Binomial likelihood to account for over-dispersion in the data (Chowell et al., 2012; Lloyd-Smith, 2007). Using the methods described in Chowell et al. (2012), we determined that the exponential rise in cases occurred up to approximately the end of November, 2015 (n = 359). Surveillance during that period by the SIVIGILA system was constant, and no unusual vector control measures were implemented during that time.

We crosschecked the robustness of the fit assumptions by fitting to the initial rise only during the month of October, and again for the month of November. If the rate of exponential rise was constant (as it should be when surveillance is constant and no unusual vector control measures are taken in the initial phase of the outbreak), the two fits should yield statistically consistent results (Towers et al., 2014).

### 2.3. Estimation of the reproduction number

ZIKV is a mosquito-borne disease for which a susceptible mosquito vector species bites an infectious human, whereupon the virus replicates in the mosquito's mid gut, and then its salivary gland cells (Chouin-Carneiro, 2016). After several days (known as the "extrinsic incubation period") (Majumder et al., 2016), ZIKV can be found in the mosquito's saliva, which then can be transmitted to other humans that the infected mosquito bites (World Health, 2016). A human, once bitten by an infected mosquito, incubates the virus for several days, whereupon they become infectious for a period of several more days (Majumder et al., 2016).

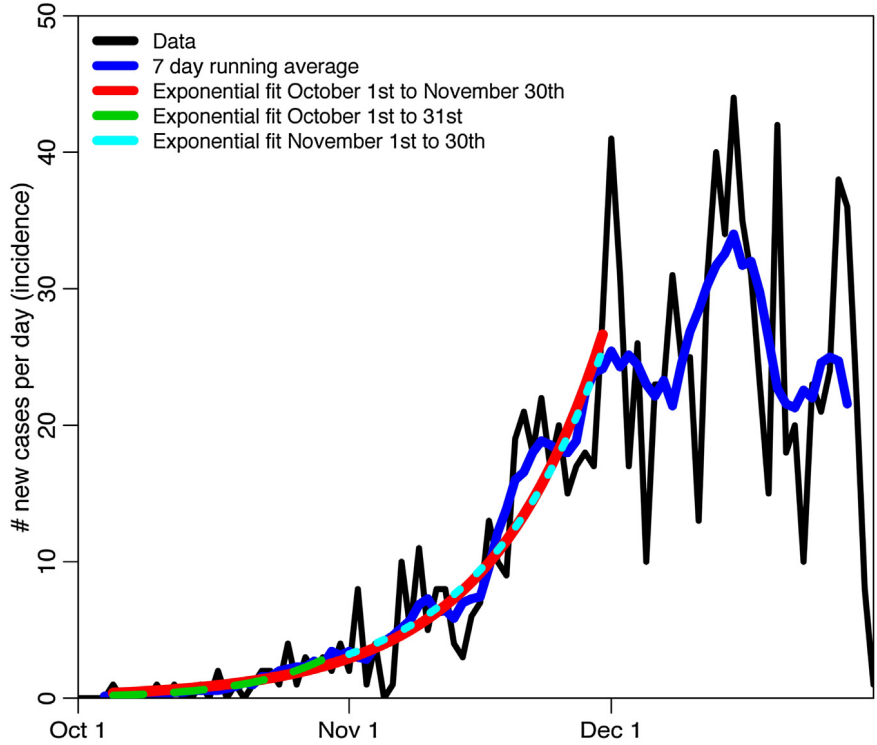

**Fig. 1.** The time series of clinically identified Zika virus disease cases in Barranquilla, Colombia, with the best-fit exponential curves to the initial rise overlaid.

In addition to these dynamics, there is evidence that ZIKV can also be sexually transmitted (Musso et al., 2015; Davidson, 2016; Dyer, 2016).

In this study, we employed a compartmental mathematical model to simulate these dynamics (Brauer et al., 2016). The model included compartments corresponding to Susceptible, Exposed, Infected, and Recovered humans, and Susceptible, Exposed, and Infected mosquitoes (and is thus known as an SEIR/SEI model).

In a population of $N_h$ humans and $N_v$ adult female mosquitoes, the susceptible adult female mosquitoes, $S_v$, upon biting an infectious human, incubate the virus for an average period, $1/\eta$, and then move to the infectious compartment, $I_v$. They die after an average of $1/\mu$ days. The transmission rate from humans to mosquitoes is $\beta_v$.

The susceptible humans, $S_h$, can be infected by being bitten by infectious mosquitoes or through direct contact with another infected human. The human then incubates the virus for an average period, $1/\kappa$ days, before becoming infectious, $I_h$. After an average of $1/\gamma$ days, the human then moves to the recovered and immune compartment, $R_h$. The transmission rate from mosquitoes to humans is $\beta$, and the direct transmission rate between humans is $\alpha$.

A similar model (without sexual transmission) was previously employed to estimate the reproduction number of DENV and CHIKV fever outbreaks (Chowell et al., 2007, 2008; Yakob and Clements, 2013), and the reproduction number of the 2013-14 outbreak of ZIKV in French Polynesia in 2007 (Kucharski, 2016).

The compartmental flow diagram for this model is shown in Fig. 2, and the system of ordinary differential equations incorporat-

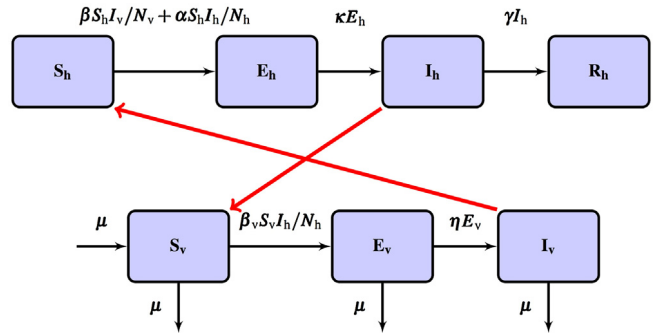

**Fig. 2.** Compartmental flow diagram of the vector-borne and sexual transmission model described in Eq. (1). Parameters are as described in Table 2.

ing the dynamics of the model were described previously in Brauer et al. (2016):

$$
\begin{aligned}
S_h^{'} &= -\beta S_h I_v / N_v - \alpha S_h I_h / N_h \\
E_h^{'} &= +\beta S_h I_v / N_v + \alpha S_h I_h / N_h - \kappa E_h \\
I_h^{'} &= \kappa E_h - \gamma I_h \\
R_h^{'} &= \gamma I_h \\
S_v^{'} &= -\beta_v S_v I_h / N_h - \mu N_v - \mu S_v \\
E_v^{'} &= +\beta_v S_v I_h / N_h - \mu + \eta\, E_v \\
I_v^{'} &= \eta E_v - \mu I_v.
\end{aligned}
\tag{1}
$$

The reproduction number of ZIKV due to vector-borne transmission alone is

$$
R0_{vector} = \beta \beta_v \frac{\eta}{\mu \gamma (\mu + \eta)},
\tag{2}
$$

and the reproduction number of ZIKV due to sexual (or other direct) transmission alone is $R0_{sex} = \alpha/\gamma$. The R0 of the model is $R0 = R0_{vector} + R0_{sex}$ (Brauer et al., 2016). Here we interpret the RHS of Eq. (2) as the reproduction number for the vector-borne component of transmission. However, it has been pointed out in the literature that the RHS of Eq. (2) can equally reasonably be interpreted as $R0^2$, rather than R0 (Brauer et al., 2016) (albeit for the purposes of this paper, it does not matter which form we use). The form in Eq. (2) is referred to as the "type reproduction number" and is the total number of secondary infections in humans originating from a human infection. In practice, we note that several past SEIR/SEI model estimates of the reproduction number for DENV, CHIKV, and ZIKV outbreaks have also used the formulation as we have it here (Chowell et al., 2007, 2008; Yakob and Clements, 2013); for consistency of comparison between our result and those estimates, we thus used the formulation as we have presented. But it must generally be kept in mind when comparing our results to those of other analyses that the explicit form of the reproduction number expression must be noted.

The expression relating the total reproduction number to the rate of exponential rise, $\rho$, and the $R0_{sex}$ is, based on the results presented in Reference (Brauer et al., 2016),

$$R0 = R0_{sex} + \left[ \frac{(\rho + \kappa)(\rho + \gamma)}{\kappa \gamma} - R0_{sex} \right] \frac{(\rho + \mu)(\rho + \mu + \eta)}{\mu(\mu + \eta)}. \quad (3)$$

Thus, given the rate of exponential rise in cases, an estimate of the R0 of sexual transmission, and estimates of the latent and infectious periods and the mosquito lifespan, we can estimate the reproduction number. Currently, there is no estimate of $R0_{sex}$, however we assume from the lack of sustained ZIKV transmission in areas free of the *Aedes* vector that $R0_{sex} < 1$.

We note here that ZIKV has a high rate of asymptomatic infection, thus the observed cases represent a small fraction of the true number (World Health, 2016; Fauci and Morens, 2016). However, this scale factor does not change the rate of exponential rise in cases; thus estimation of the rate of exponential rise with only partial observation of cases is a robust means to estimate the reproduction number.

To estimate the probability distribution for R0, given probability distributions for $R0_{sex}$, $\rho$, $\kappa$, $\gamma$, $\eta$, and $\mu$, we performed ten thousand Monte Carlo iterations, whereby we randomly sampled values of $\rho$ and the model parameters from their probability distributions, and calculated the resulting estimate of R0 at each iteration using Eq. (3), and also the fraction of cases due to sexual transmission estimated from the numerical solution to Eq. (1), both for the entire simulated epidemic curve, and also the exponential rise region, identified using the methods of Reference (Chowell et al., 2012). The serial interval of ZIKV is the sum of the incubation periods plus half the infectious periods, $T = 1/\kappa + 1/\eta + 0.5/\mu + 0.5/\gamma$, as described in Reference (Svensson, 2007); we thus used the constraint that the serial interval derived from the sampled parameters had to be within the observed serial interval of 10 to 23 days (Majumder et al., 2016). The resulting distribution of the estimates of R0 formed the estimate of the probability density for that quantity (Gardner and O'Neill, 1983).

We currently know that $R0_{sex} < 1$, but do not know its value, thus the probability distribution used for $R0_{sex}$ was Uniform(0,1). The probability distributions for $\kappa$, $\gamma$, $\eta$, and $\mu$ were derived from the ranges of the literature estimates for these quantities, assuming a Uniform probability distribution over the range. The estimates used to assess the range of the Uniform probability distributions for these parameters are shown in Table 2. The probability distribution for the rate of exponential rise, $\rho$, was derived from the likelihood fit to the data, and was Normally distributed about the best-fit estimate.

## 3. Results

### 3.1. Estimation of the rate of initial exponential rise in cases

The results of the fit of an exponential rise to the daily incidence data of clinically identified ZIKV cases in Barranquilla from the beginning of October up to the end of November (n = 359) are shown overlaid on the data in Fig. 1. The estimated rate of exponential rise was $\rho = 0.076$ days$^{-1}$, with 95% confidence interval [0.066, 0.087] days$^{-1}$.

We crosschecked our assumption of a constant rate of exponential rise in the initial data by fitting only up to the end of October (n = 32), yielding $\rho = 0.111$ [0.060, 0.167] days$^{-1}$. We additionally fit to the data from the beginning to the end of November (n = 327), yielding $\rho = 0.071$ [0.055, 0.089] days$^{-1}$. These two rates of rise were statistically consistent ($\chi^2_{\nu = 1}$ test p = 0.15), thus the exponential rate of rise in ZIKV cases appeared to be consistent over the entire period from the beginning of October to the end of November. These two fits are also shown overlaid on the incidence data in Fig. 1.

Excluding pregnant women from the fit only excluded three cases up to the end of November, and resulted in negligible changes to the estimated exponential rise.

### 3.2. Estimation of the reproduction number

With the use of Eq. (1), we obtained the average value of the estimated reproduction number as R0 = 3.8 with [2.4,5.6] 95% CI, and the fraction of cases due to sexual transmission was 0.23 [0.01,0.47], with 95% confidence. The latter quantity was estimated from the entire simulated epidemic curve. In the exponential rise region only, the fraction of cases was 0.27 [0.01,0.56], with 95% confidence

## 4. Discussion

We have estimated the R0 of a ZIKV outbreak occurring in an area typical of the regions in the Americas that are currently affected by the disease.

Despite the pressing need to estimate the R0 for a newly emerging pandemic disease, there are currently only two other estimates of the R0 of a ZIKV disease outbreak of which the authors are aware; the R0 of the 2013-14 ZIKV outbreak in French Polynesia was recently estimated to be between 2.6 to 4.8 (Kucharski, 2016), in statistical agreement with the estimate of the R0 obtained by this analysis, R0 = 3.8 [2.4, 5.6]. The R0 of ZIKV based on data from Colombia, Brazil, and El Salvador at the beginning of the 2015/2016 outbreak was estimated to be R0 = 2.055 [0.523,6.300] with an estimated percentage of cases due to sexual transmission being 3.044% [0.123%,45.73%] (Gao et al., 2016). This estimated percentage of cases due to sexual transmission is in statistical agreement with our estimate, that the percentage was 23 [1,47]% with 95% confidence. Note that the vector-borne R0 used in Reference (Gao et al., 2016) is the square root of the type-reproduction number we use, thus the results in Reference (Gao et al., 2016) for R0 must be approximately squared for comparison to those of this analysis. However, due to the very large confidence interval on R0 estimated by Reference (Gao et al., 2016), both the squared and un-squared values are in statistical agreement with our estimate. We note here, as an aside, that the confidence interval for R0 estimated by Reference (Gao et al., 2016) paradoxically includes values less than one, even though the estimates were obtained by fitting to data from three large epidemics, thus values of R0 less than 1 should have been impossible.

SEIR/SEI model estimates for the R0 of DENV outbreaks generally fall between approximately 1.5 to 3 (see, for instance, References (Chowell et al., 2007, 2008)), and around 4 for CHIKV outbreaks (Yakob and Clements, 2013; Svensson, 2007). The estimated R0 for the ZIKV outbreak in Barranquilla is similar to that of CHIKV outbreaks, but is larger than the estimates for DENV outbreaks, despite the fact that the vector species is the same for the diseases. This could be due to a variety of reasons, including a sexual transmission component to ZIKV (Fauci and Morens, 2016; Musso et al., 2015), and/or partial prior immunity to DENV in hyperendemic areas, and/or shorter incubation periods for ZIKV in the vectors, and/or a longer infectious period in humans, or even increased susceptibility to infection.

**Table 2**
Model parameters used in this analysis. The incubation and infectious periods, and the mosquito average lifespan, were determined from the cited references, and the references therein.

| Parameter | Definition | Estimate (days) | Reference |
|---|---|---|---|
| $1/\kappa$ | Intrinsic (human) latent period | 3–12 | (Majumder et al., 2016) |
| | | 2–7 | (World Health, 2016) (Kucharski, 2016) |
| | | 2–6 | |
| | Assume Uniform (2,12) | | |
| $1/\gamma$ | Human infectious Period | 3–5 | (Majumder et al., 2016) |
| | | 3–7 | (Kucharski, 2016) |
| | Assume Uniform (3,7) | | |
| $1/\eta$ | Extrinsic (mosquito) infectious period | 4–6 | (Majumder et al., 2016) |
| | | 10–15 | (Hayes, 2009) (Kucharski, 2016) |
| | | 8–13 | |
| | Assume Uniform (4,15) | | |
| $1/\mu$ | Average mosquito lifespan | 6–15 | (Chowell et al., 2008, 2007) |
| | | 10–20 | (Yakob and Clements, 2013) (Kucharski, 2016) |
| | | 6–10 | |
| | Assume Uniform (6,20) | | |
| $1/\eta + 1/\kappa + \frac{1}{2}\gamma + \frac{1}{2}\mu$ | Serial Interval | 10–23 | (Majumder et al., 2016) |

It is important to note here that our estimate of the ZIKV R0 is model dependent, as indeed are all such estimates in the literature. However, for ease of comparison of results we have used a similar model to that employed to analyze the French Polynesia ZIKV outbreak (Kucharski, 2016), DENV outbreaks (Chowell et al., 2007, 2008), and a CHIKV outbreak (Yakob and Clements, 2013). While we have not explored other model formulations in this analysis, it would be easy to use our estimate of the exponential rise with other compartmental model formulations, and/or other model parameter values, in order to extract alternative estimates of R0.

Our model assumed homogeneity in contacts between members of the population, similar to other studies (Gao et al., 2016). In reality, contacts between members in the population, especially sexual contacts, are generally highly networked and gender dependent. However, early in an epidemic during the exponential growth phase, the rate of exponential growth is driven largely by the R0 and serial interval of the disease (Barthélemy et al., 2005; Brauer, 2012), and it is only later in the epidemic that model mixing assumptions affect forecasts of the outbreak. Thus, despite the simplified assumptions of mixing in our model we can estimate the R0, but the model should not be used to estimate the final size of the epidemic, or to quantify the efficacy of control strategies, including those aimed at sexual transmission.

There is currently much uncertainty related to the epidemiological parameters of ZIKV infection, such as the human and mosquito latent periods, and R0 of sexual transmission, which are almost wholly responsible for the somewhat broad confidence interval on our estimate of R0. Further study will help to constrain these parameters, and allow for more precise estimates of the R0. Again, because we have provided our estimate of the exponential rise in cases for this outbreak, it would be easy in the future to re-calculate the R0 once these parameters are better known.

Our study is based on clinical syndromic surveillance data, similar to many other studies of arboviral outbreaks (for instance, References (Cardona-Ospina et al., 2015; Chowell et al., 2007, 2008)), due to limited laboratory testing resources in developing countries where many of these outbreaks occur. ZIKV has similar symptoms to some of the symptoms seen in mild cases of DENV (Brasil, 2016), and further study is needed in the future to determine how potential misdiagnosis may affect apparent temporal dynamics in places where other arboviral diseases may be co-circulating (Fauci and Morens, 2016). However, it must be pointed out that initial studies indicate very low frequency of contamination by other arboviral infections in cases identified by ZIKV clinical surveillance in areas where DENV and CHIKV co-circulate (Brasil, 2016; Tognarelli et al., 2015).

Changes in human behavioral responses, including health-seeking behaviors, are always a potential confounder in quantitative studies of outbreaks of high profile newly emerging diseases when fear may be incited in the population (see, for example, References (Fenichel et al., 2011; Lau et al., 2005; Plucinski et al., 2015)). However, the data we consider in this analysis in October and November 2015 are early in the outbreak, largely before significant international attention began to be paid to ZIKV and the birth defects and other disorders it can cause; we note here that the first epidemiological alert by the WHO and the Pan American Health Organisation detailing a possible link between ZIKV infection and microcephaly was on November 17, 2015.[1] After the period considered during this study, there was a sharp increase in the number of cases identified in pregnant women, potentially indicating increasing concern in this group as time went on; during the month of December, 39 of the 782 (5%) identified cases with dates of initial symptoms were pregnant females, which is not statistically consistent with the 3 of the 359 (1%) cases considered during October and November ($\chi^2_{\nu=1}$ p < 0.001). In addition, the fraction of cases involving pregnant women during October and November 2015 was statistically consistent with the crude birth rate in Colombia (16/1000),[2] indicating no apparent heightened concern in that group during that period.

Sexual transmission has been noted to play a role in ZIKV transmission (World Health, 2015), but the relative contribution of sexual transmission cannot be determined based on exponential rise in incidence alone unless further information is known. In this case, we use the fact that sustained transmission in areas without *Aedes* mosquitoes has not been observed, thus the R0 of sexual or

---

other direct transmission must be less than 1. Our estimated percentage of cases due to sexual transmission when *Aedes* mosquitoes are also present may be as high as 47%, with 95% confidence. As such, safe sexual practices may significantly reduce incidence.

## 5. Conclusions

ZIKV has spread explosively in the Americas in recent months. Here we employed an SEIR/SEI mathematical model of the spread of vector-borne disease with additional sexual transmission to estimate the basic reproduction number, R0, of the ZIKV outbreak that began in October 2015, in Barranquilla, Colombia. We estimated the R0 to be R0 = 3.8 with 95% CI [2.4,5.6], and a one standard deviation uncertainty of 0.8. This is among the first estimates of the R0 of a ZIKV outbreak in the Americas, and is important to assessment of outbreak risk in new areas. Our modeling analysis estimates that up to 47% of ZIKV cases in Barranquilla, with 95% confidence, may have been due to sexual contact alone. Thus safe sexual practices may significantly reduce incidence during ZIKV outbreaks in tropical and semi-tropical areas.

## Acknowledgments

The authors are very grateful to Dr. Elsa Bravo de Plata and Dr. Pedro Arango of the Secretaria Distrital de Salud de Barranquilla, Colombia, for their advice and assistance related to these studies.

The authors are grateful for the following sources of funding that supported these studies:

FB: Natural Sciences and Engineering Research Council of Canada, Grant OGPIN 203901-99;

AKIF, CMERV: Proyecto de Regalias, Convenio0103-2015-000039, Actividad MGA [955869];

CCC, ST, AM: National Science Foundation (DMS-1263374, DUE-1101782), National Security Agency (H98230-14-1-0157), the Office of the President of ASU, and the Office of the Provost of ASU.